# Photoemission of the Upconverted Hot Electrons in Mn-doped CsPbBr$_3$ Nanocrystals

Chih-Wei Wang[1], Xiaohan Liu[2], Tian Qiao[1], Mohit Khurana[2], Alexey V. Akimov[2], Dong Hee Son[1,3]

[1]Department of Chemistry, Texas A&M University, College Station, Texas 77843, USA
[2]Department of Physics, Texas A&M University, College Station, Texas 77843, USA
[3]Center for Nanomedicine, Institute for Basic Science and Graduate Program of Nano Biomedical Engineering, Advanced Science Institute, Yonsei University, Seoul 03722, Republic of Korea

**Abstract**

Hot electrons play a crucial role in enhancing the efficiency of photon-to-current conversion or photocatalytic reactions. In semiconductor nanocrystals, energetic hot electrons capable of photoemission can be generated via the upconversion process involving the dopant-originated intermediate state, currently known only in Mn-doped cadmium chalcogenide quantum dots. Here, we report that Mn-doped CsPbBr$_3$ nanocrystals are an excellent platform for generating hot electrons via upconversion that can benefit from various desirable exciton properties and the structural diversity of metal halide perovskites (MHP). 2-dimensional Mn-doped CsPbBr$_3$ nanoplatelets are particularly advantageous for hot electron upconversion due to the strong exciton-dopant interaction mediating the upconversion process. Furthermore, nanoplatelets reveal evidence for the hot electron upconversion via long-lived dark exciton in addition to bright exciton that may enhance the upconversion efficiency. This study not only establishes the feasibility of hot electron upconversion in MHP host but also demonstrates the potential merits of 2-dimensional MHP nanocrystals in hot electron upconversion.

**Main**

Hot electrons from metallic or semiconductor nanostructures have been investigated extensively due to their ability to drive energetically and kinetically challenging electron transfer in chemical processes or in devices.[1, 2, 3, 4, 5] In semiconductor nanocrystals, exciton-to-hot electron 'upconversion' is known as a unique pathway producing energetic hot electrons from weak visible excitation, so far observed only in Mn-doped cadmium chalcogenide quantum dots (QDs).[6, 7, 8] The hot electron upconversion combines the energy of two excitations to produce a higher-energy electron through a long-lived intermediate state from the doped $Mn^{2+}$ ions mediated by the exciton-dopant exchange interaction. It bears a similarity to the photon upconversion producing the higher-energy photons in its upconverting functionality,[9, 10, 11, 12] while the photophysical pathways and the final states are very different. Recently, doping of $Mn^{2+}$ was successfully performed in metal halide perovskite (MHP) nanocrystals, which are emerging as the superior alternative to many other semiconductor nanocrystals as the source of photons, charges and transferable energy.[13, 14, 15, 16, 17, 18] Various Mn-doped MHP nanocrystals with different degree of quantum confinement and dimensionality, such as the QDs and nanoplatelets (NPLs), were synthesized with the progress of doping stratagies,[19, 20, 21, 22] and the sensitization of $Mn^{2+}$ by exciton was also confirmed.[23, 24, 25, 26] This suggests the possible hot electron upconversion that can take advantage of the excellent exciton properties and the structural diversity known in MHP nanocrystals, while the understanding on the exciton-dopant interaction in these materials is still immature.

Here, we show that hot electron upconversion is indeed possible in Mn-doped MHP nanocrystals using quantum-confined $CsPbBr_3$ QDs and NPLs as the host nanocrystals, from which the hot electrons emitted above the vacuum level were directly detected. In particular, 2-dimensional Mn-doped $CsPbBr_3$ NPLs exhibited significant advantages compared to the QDs due to the stronger exciton-dopant interaction mediating the upconversion process. Furthermore, Mn-doped $CsPbBr_3$ NPLs revealed evidence for the hot electron upconversion by dark exciton in addition to bright exciton, which can potentially benefit from the longevity of dark exciton. In this study, we established the feasibility of hot electron upconversion in this new family of semiconductor nanocrystals and show the potential advantages of 2-dimensional MHP nanocrystals in hot electrons upconversion.

**Results and Discussion**

Generation of hot electrons via upconversion in Mn-doped semiconductor nanocrystals requires two sequential processes, where the long-lived (~ms) excited ligand field state of $Mn^{2+}$ ions, $|Mn^{2+}\rangle^*$, functions as the intermediate state for the upconversion.[27] Fig. 1a illustrates the two processes, i.e., the sensitization of $Mn^{2+}$ by exciton to populate $|Mn^{2+}\rangle^*$, and the subsequent Auger energy transfer from $|Mn^{2+}\rangle^*$ to another exciton to produce the hot electron. The hot electrons generated via upconversion process gain large excess energy from $|Mn^{2+}\rangle^*$ (e.g., >2 eV) above the conduction bandedge, therefore are capable of performing long-range transfer over a high barrier and even generating the photoelectron emission.[6, 28]

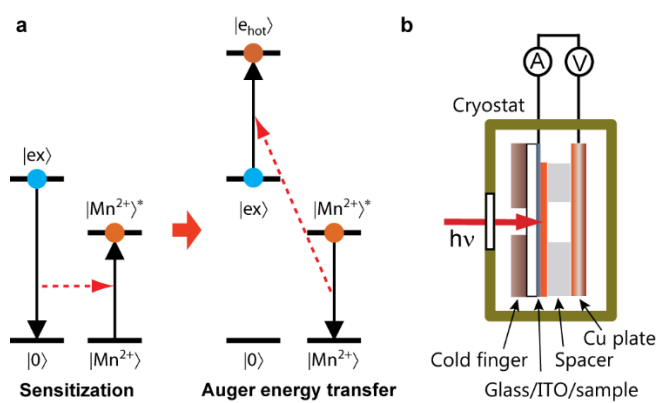

**Fig. 1: Hot electron upconversion and hot electron photoemission current measurement. A**, Photophysical pathways of hot electron upconversion. $|0\rangle$ and $|ex\rangle$ are ground and exciton state of the host nanocrystals. $|Mn^{2+}\rangle$ and $|Mn^{2+}\rangle^*$ are ground and excited state of doped $Mn^{2+}$. $|e_{hot}\rangle$ represents the state generating hot electrons. **B**, Schematic diagram of the photoemission measurement setup, Ⓐ and Ⓥ represent the ammeter with variable bias.

In this study, we chose cesium lead halide ($CsPbX_3$) nanocrystals as the candidate host for hot electron upconversion, for which $Mn^{2+}$ doping is relatively well established. Among $CsPbX_3$ (X=Cl, Br, I), only $CsPbCl_3$ has the sufficiently high bulk bandgap (3.04 eV) that allows the sensitization of $Mn^{2+}$ without quantum confinement in the host, while ultraviolet light is required for efficient photoexcitation. For $CsPbBr_3$ with the lower bulk bandgap (2.35 eV), imposing quantum confinement increases the bandgap[29] and enables the sensitization of $Mn^{2+}$ via visible excitation. The same strategy is less effective for $CsPbI_3$ since its bulk bandgap is even lower (1.67 eV).[24] Here, we focus on Mn-doped $CsPbBr_3$ QDs (6 nm, ~10 % doping) and NPLs (~2 nm thick, 30-50 nm lateral dimension, ~10 % doping) representing two common morphologies of MHP nanocrystals that also impose different degree of quantum confinement. The details of the synthesis, the structural characterization, and the summary of the optical properties (Table S1) are in Supplementary Information.

Hot electron upconversion in Mn-doped $CsPbBr_3$ QDs and NPLs was studied by directly measuring the photocurrent of hot electrons emitted above the vacuum level under *cw* excitation at 405 nm. Fig. 1b shows the schematics of the experimental setup, where the Mn-doped nanocrystals spin-cast on an indium tin oxide (ITO)-coated glass substrate functions as the photocathode generating hot electrons. A copper plate placed 2 mm away from the photocathode collects the photoemitted hot electrons from the nanocrystals. The current density of the photoemitted hot electrons ($J_{hot}$) was measured under the vacuum at varying temperatures, excitation intensities and electrical biases between the photocathode and copper electrode as described in Methods and Supplementary Information. The average energy available from $|Mn^{2+}\rangle^*$ is ~2 eV or larger and the quantum-confined electron level in $CsPbBr_3$ nanocrystals is <3 eV below the vacuum level based on the bulk band edge level.[30] Therefore, the measured $J_{hot}$ accounts for a subpopulation of the hot electrons that is above the vacuum level. While the energy spectrum of these hot electrons is unknown, the measurement of $J_{hot}$ is the most direct confirmation of the hot electron upconversion. Furthermore, it allows for the comparison of the relative 'hotness' of hot electrons generated from Mn-doped $CsPbBr_3$ QDs and NPLs having different bandgap via bias-dependent $J_{hot}$ as will be discussed later.

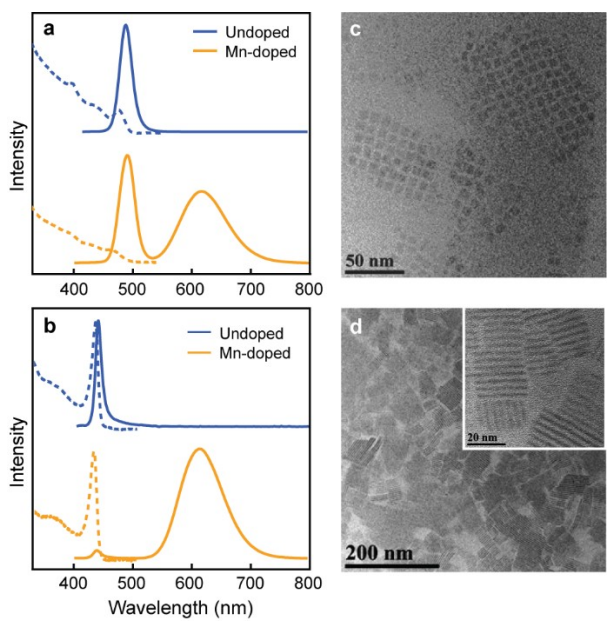

**Fig. 2: Absorption and photoluminescence spectra and TEM images of Mn-doped CsPbBr$_3$ QDs and NPLs. a-b**, Solution phase absorption (dashed) and photoluminescence spectra (solid) of Mn-doped MNP CsPbBr$_3$ QDs (**a**) and NPLs (**b**) compared with their undoped counterparts. **c-d**, TEM images of Mn-doped CsPbBr$_3$ QDs (**c**) and NPLs (**d**).

Fig. 2a and b show the absorption and photoluminescence (PL) spectra of Mn-doped CsPbBr$_3$ QDs and NPLs at the ambient temperature compared with those of undoped counterparts. The representative transmission electron microscope (TEM) images are shown in Fig. 2c and d. They exhibit the strong sensitized PL from *d-d* transition of Mn$^{2+}$ near 610 nm in addition to the weaker exciton PL appearing at 490 nm and 440 nm respectively for QDs and NPLs. The relative intensity of Mn$^{2+}$ PL with respect to exciton PL is stronger in NPLs than in QDs, which reflects the difference in the rate of sensitization that depends on the degree of quantum confinement as well as doping concentration. The rate of sensitization per each Mn$^{2+}$ dopant is considered more than an order of magnitude faster in NPLs than in QDs (See Supplementary Information),[21, 25, 26] which is responsible for the large differences in their hot electron upconversion properties under varying conditions as discussed below.

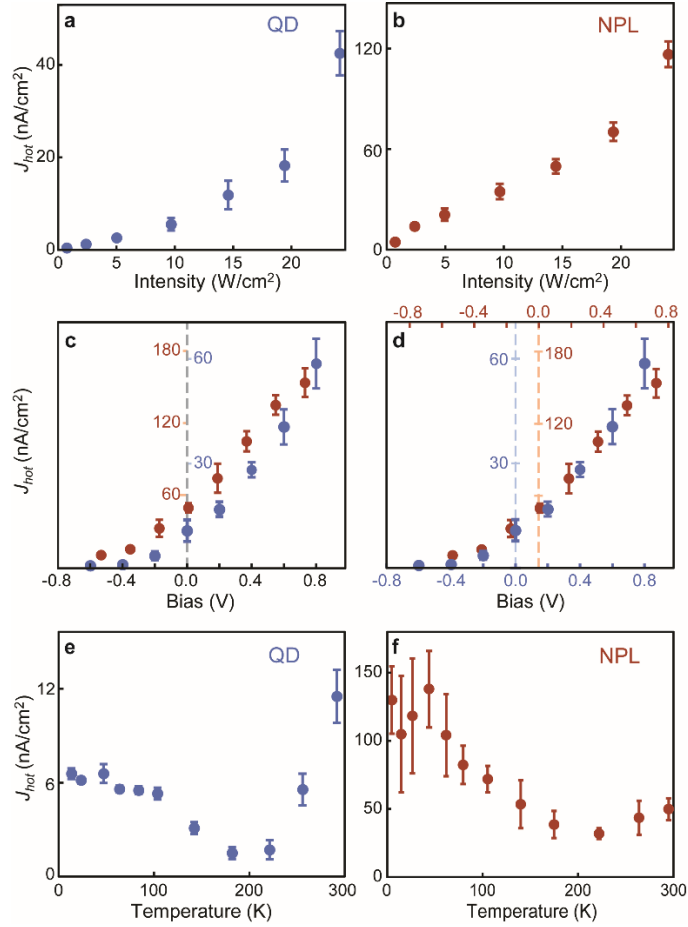

**Fig. 3**: **Excitation intensity, bias and temperature dependence of hot electron photocurrent density from Mn-doped CsPbBr$_3$ QDs and NPLs. a-b**, Excitation intensity-dependent hot electron photoemission current density ($J_{hot}$) of Mn-doped CsPbBr$_3$ QDs (**a**) and NPLs (**b**). **c**, Comparison of the bias-dependent $J_{hot}$ of Mn-doped CsPbBr$_3$ QDs (blue) and NPLs (red). **d**, Comparison of $J_{hot}$ of Mn-doped CsPbBr$_3$ QDs (blue) and NPLs (red) with 0.15 V relative shift of bias on the x-axis. **e-f**, Temperature-dependent $J_{hot}$ of Mn-doped CsPbBr$_3$ QDs (**e**) and NPLs (**f**).

Fig. 3a and b show the excitation intensity-dependent $J_{hot}$ of Mn-doped CsPbBr$_3$ QDs and NPLs at the ambient temperature and 0 V bias between the two electrodes. $J_{hot}$ from the undoped QDs and NPLs were negligible (<0.1 nA/cm$^2$). The absolute values of $J_{hot}$ cannot be directly compared between the QDs and NPLs because of the difference in the amount the nanocrystals deposited on the electrode which is difficult to quantify accurately. However, each nanocrystal sample exhibited reproducible behavior in all $J_{hot}$ measurements. Since the hot electron upconversion is the sequential biphotonic process, one would expect $J_{hot}$ quadratic to the excitation intensity when far from saturating the population of the intermediate state, $|Mn^{2+}\rangle^*$. $J_{hot}$ of Mn-doped CsPbBr$_3$ QDs exhibits a superlinear increase of $J_{hot}$ with log-log slope of ~1.7 when the lowest two data points are excluded. (Fig. S1a) On the other hand, Mn-doped CsPbBr$_3$ NPLs show a more linear increase of $J_{hot}$ with log-log slope close to 1 (Fig. S1b), which may indicate the earlier saturation of $|Mn^{2+}\rangle^*$ population leading to the quasi-linear increase of $J_{hot}$. In order to examine such possibility, the approximate excitation intensity-dependent steady state population of $|Mn^{2+}\rangle^*$ estimated for Mn-doped CsPbBr$_3$ QDs and NPLs are compared (Fig. S2). The details of the kinetic model and the assumptions made in the comparison are in Supplementary Information. The steady state $|Mn^{2+}\rangle^*$ population increased continuously with increasing excitation intensity in Mn-doped CsPbBr$_3$ QDs, whereas it saturated rapidly in Mn-doped CsPbBr$_3$ NPLs, qualitatively consistent with the observed difference in the intensity-dependent $J_{hot}$. A crucial factor responsible for this difference between the QDs and NPLs is the sensitization rate that differ by an order of magnitude. (~1/1000 ps$^{-1}$ for the QDs and ~1/50 ps$^{-1}$ for the NPLs from Supplementary Information) The saturation of $|Mn^{2+}\rangle^*$ at the lower excitation intensity benefits hot electron upconversion since the photon-to-hot electron conversion ratio should also saturate at the lower intensity.

Fig. 3c compares the bias-dependent $J_{hot}$ of Mn-doped CsPbBr$_3$ QDs and NPLs at the ambient temperature and excitation intensity of 15 W/cm$^2$. The increase of $J_{hot}$ with increasing positive bias producing a weak electric field (<4V/cm) is considered partially due to the modified trajectory of the photoemitted hot electrons that increases the number of hot electrons collected by the copper electrode.[6] At negative biases, $J_{hot}$ decreases with more negative bias until it reaches the stopping voltage of the photoemitted hot electrons. Therefore, the decay of $J_{hot}$ at negative bias can provide some information on the excess kinetic energy of the photoemitted hot electrons. It is noteworthy that $J_{hot}$ vs bias curve of Mn-doped CsPbBr$_3$ QDs appears shifted by ~0.15 V with respect to that of Mn-doped CsPbBr$_3$ NPLs. This can be seen more clearly in Fig. 3d comparing the two $J_{hot}$ vs bias curves after adding 0.15 V relative shift on the horizontal axis. It can be interpreted as the higher average kinetic energy of the photoemitted hot electrons from NPLs than from QDs by 0.15 eV assuming that the lineshape of $J_{hot}$ vs bias is similar between the two nanocrystals. The additional 0.15 eV of hot electron energy is consistent with ~0.28 eV larger bandgap of Mn-doped CsPbBr$_3$ NPLs than that of Mn-doped CsPbBr$_3$ QDs. Considering the similar effective mass of electron and hole in CsPbBr$_3$,[31] the additional elevation of the confined electron level in NPLs reflected in hot electron energy should be approximately half of the difference in the bandgap.

Fig. 3e and f compare the temperature-dependent $J_{hot}$ of Mn-doped CsPbBr$_3$ QDs and NPLs at the excitation intensity of 15 W/cm$^2$ and 0 V bias between the two electrodes. $J_{hot}$ of Mn-doped CsPbBr$_3$ QDs show a prominent dip at ~200 K, whereas Mn-doped CsPbBr$_3$ NPLs show a more monotonous increase of $J_{hot}$ with decreasing temperature. The difference in the temperature-dependent $J_{hot}$ was observed reproducibly in multiple batches of QDs and NPLs, indicating the clear correlation to the structural details of the nanocrystals. The temperature-dependent $J_{hot}$ reflects the temperature-dependent competition among many dynamic processes including the relaxation of exciton and $|Mn^{2+}\rangle^*$, sensitization of $Mn^{2+}$, Auger energy transfer, and trapping/detrapping of exciton. Since the rates of these processes depend on factors, such as the doping density and structure, energetics of defects and degree of quantum confinement,[20, 32, 33] different temperature dependence of $J_{hot}$ in different nanocrystals is not unexpected. Presently, a complete picture of the temperature-dependent competitive dynamics that can explain the entire temperature-dependent $J_{hot}$ is not available due to the insufficient knowledge on the rates of all the competing processes. Nevertheless, the temperature-dependent steady state PL can provide an insight into the contribution of the first step of the upconversion, i.e., sensitization of $Mn^{2+}$. Furthermore, the role of dark exciton on the sensitization and hot electron upconversion can be inferred from the temperature-dependent PL data and $J_{hot}$ as will be discussed shortly.

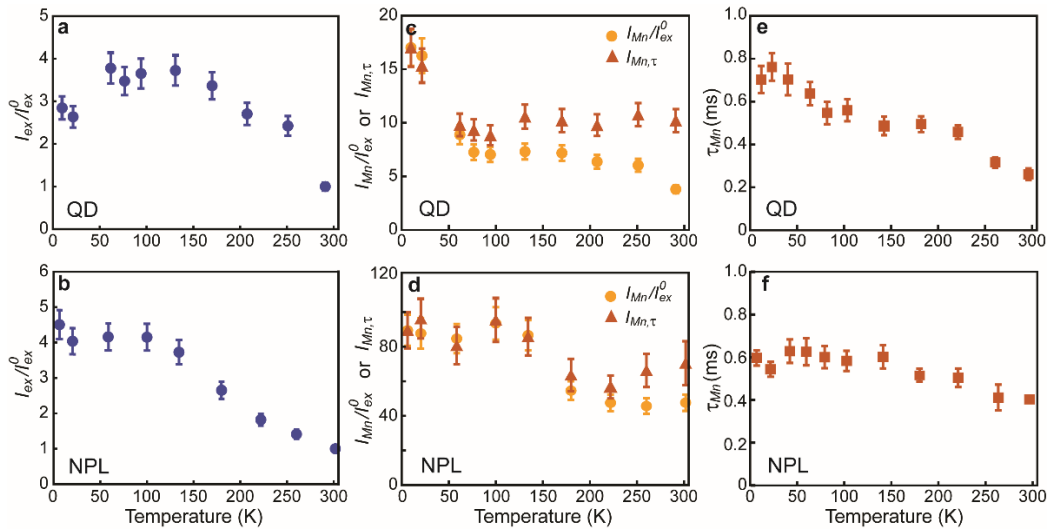

**Fig. 4: Temperature-dependent exciton and Mn$^{2+}$ PL intensities and Mn$^{2+}$ PL lifetime of Mn-doped CsPbBr$_3$ QDs and NPLs. a-b**, Temperature dependence of exciton PL intensity of Mn-doped CsPbBr$_3$ QDs (**a**) and NPL (**b**). **c-d**, Temperature dependence of Mn$^{2+}$ PL intensity (orange) and $I_{Mn,\tau}$ (red) of Mn-doped CsPbBr$_3$ QDs (**c**) and NPLs (**d**). PL intensities in panel **a-d** are

normalized to the exciton PL intensity at the ambient temperature. **e-f**, $Mn^{2+}$ PL lifetime of Mn-doped $CsPbBr_3$ QDs (**e**) and NPLs (**f**).

Fig. 4a-d compare the temperature-dependent PL intensity of exciton ($I_{ex}$) and $Mn^{2+}$ ($I_{Mn}$) normalized to exciton PL intensity at the ambient temperature ($I_{ex}^0$) for Mn-doped $CsPbBr_3$ QDs and NPLs under the condition that excites much less than 1 exciton/nanocrystal during the lifetime of $|Mn^{2+}\rangle^*$. Temperature-dependent $|Mn^{2+}\rangle^*$ lifetime ($\tau_{Mn}$) is also shown in Fig. 4e and f. The full temperature-dependent PL spectra of Mn-doped $CsPbBr_3$ QDs and NPLs and $Mn^{2+}$ PL decay curves are in Fig. S5-6. For both Mn-doped nanocrystals, the PL is dominated by $Mn^{2+}$ PL at all temperatures as indicated by the large values of $I_{Mn}/I_{ex}^0$ in Fig. 4c-d. Since $I_{Mn}$ is proportional to the product of the sensitization efficiency and the quantum yield of $Mn^{2+}$ PL, $\tau_{Mn}$-normalized $I_{Mn}$ ($I_{Mn,\tau}=aI_{Mn}/\tau_{Mn}$, $a$=constant) represents the relative sensitization efficiency that accounts for the nonradiative decay of $|Mn^{2+}\rangle^*$. (See Supplementary Information) In Fig. 4c-d, the temperature dependence of $I_{Mn,\tau}$ (red) is overlayed on top of $I_{Mn}/I_{ex}^0$ (orange), where $a$ is chosen to match $I_{Mn,\tau}$ and $I_{Mn}/I_{ex}^0$ at 10 K. Since very low excitation intensity that does not allow the Auger energy transfer is used in the PL measurement, $I_{Mn,\tau}$ reflects the relative efficiency of sensitization between single exciton and all $Mn^{2+}$ ions the exciton can reach in each nanocrystal. While the hot electron upconversion occurs at the higher excitation intensities with a substantial steady state $|Mn^{2+}\rangle^*$ population, $I_{Mn,\tau}$ can still be used to compare the 'relative' change in sensitization efficiency with the change of temperature at a given excitation intensity.

$I_{Mn,\tau}$ in Fig. 4c-d shows 30-50 % increases as the temperature decreases below ~20 K for Mn-doped $CsPbBr_3$ QDs and ~100 K for Mn-doped $CsPbBr_3$ NPLs compared to the higher temperatures, indicating more efficient sensitization of $Mn^{2+}$. This can in principle be interpreted as the decreasing nonradiative decay of exciton competing with both the sensitization and radiative exciton recombination with decreasing temperature,[24, 32] which is also consistent with generally increasing $I_{ex}/I_{ex}^0$ with decreasing temperature. On the other hand, dipole-forbidden dark exciton is reported to be the dominant emitting state in strongly confined $CsPbBr_3$ QDs and NPLs at sufficiently low temperatures suppressing the thermal equilibration between the bright and dark states.[34, 35, 36] Therefore, the interpretation of $I_{Mn,\tau}$ at low temperatures is more complex and should consider the possible dark exciton-mediated sensitization of $Mn^{2+}$.

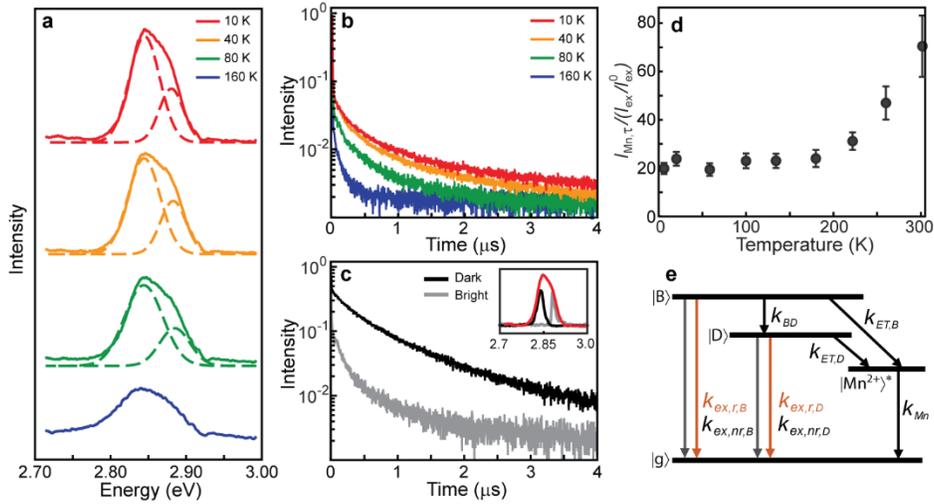

**Fig. 5: Dark exciton-mediated hot electron upconversion in Mn-doped CsPbBr$_3$ NPLs. a**, Temperature-dependent exciton PL spectra of Mn-doped CsPbBr$_3$ NPLs. The PL spectra at 10-80 K are fit to two Gaussian peaks, one corresponding to dark exciton (lower energy) and the other to bright exciton (higher energy). The bright-dark energy splitting (ΔE$_{BD}$) is 35 meV. **B**, Exciton PL intensity decay curves at different temperatures. **C**, Wavelength-selected PL decay at 10 K. Black and grey curves were obtained with the filter spectra overlayed with the PL spectra shown in the inset, which detects primarily dark and bright exciton PL. **d**, Temperature-dependent $I_{Mn,\tau}/(I_{ex}/I_{ex}^0)$ representing $k_{ET}/k_{ex,r}$. **E**, Kinetic scheme of sensitization from bright and dark exciton. $K_{ex,r,B(D)}$ and $k_{ex,nr,B(D)}$ are the radiative and nonradiative decay rate of bright (dark) exciton. $K_{ET,B(D)}$ is the sensitization rate from bright (dark) exciton. $K_{BD}$ is the transition rate from bright to dark state. $K_{Mn}$ is the decay rate of $|Mn^{2+}\rangle^*$.

An important insight into the role of dark exciton in sensitization can be obtained from Mn-doped CsPbBr$_3$ NPLs, in which the dark exciton PL can be observed in a large temperature range. In undoped CsPbBr$_3$ nanocrystals, NPLs with strongly confined 2-dimensional exciton exhibit the larger bright-dark energy splitting (ΔE$_{BD}$) compared to the QDs,[35, 36] which also give the access to the dark exciton in the wider temperature range. Interestingly, Mn$^{2+}$ doping in 2 nm-thick CsPbBr$_3$ NPLs increases ΔE$_{BD}$ from 22 meV to 35 meV accompanied by the wider temperature range dark exciton PL is dominant up to ~100 K. Dark exciton PL with ~1 μs lifetime can be seen in the temperature-dependent exciton PL spectra and decay curves (Fig. 5a, b) and the wavelength-selected exciton PL decay curves (Fig. 5c). The lifetime shortening of the μs-component of the PL via mixing of the bright and dark states under external magnetic field, further supporting its dark exciton origin, is shown in Fig. S5. At 160 K, μs-component of the PL is minimal indicating that sensitization above this temperature is predominantly from bright exciton.

Therefore, the examination of $Mn^{2+}$ PL intensities between the two temperature regions, where the nature of the populated exciton state is very different, can provide a clue to the role of dark exciton in sensitization. For this purpose, we use $I_{Mn,\tau}/(I_{ex}/I_{ex}^0)$ that is proportional to $k_{ET}/k_{ex,r}$ in a simple kinetic picture assuming the exciton population dissipation via radiative ($k_{ex,r}$) and nonradiative ($k_{ex,nr}$) exciton recombination and the sensitization of $Mn^{2+}$ ($k_{ET}$). (See Supplementary Information) The temperature-dependent $I_{Mn,\tau}/(I_{ex}/I_{ex}^0)$ of Mn-doped $CsPbBr_3$ NPLs shown in Fig. 5d initially decreases with cooling, but becomes nearly constant below 170 K. While the quantitative understanding of the full temperature dependence of $k_{ET}/k_{ex,r}$ is rather a complex task, we focus our attention on 170-10 K range exhibiting nearly constant $k_{ET}/k_{ex,r}$ despite the change in the dominant emitting state from bright to dark exciton. The bright-dark transition time in undoped 2 nm-thick $CsPbBr_3$ NPLs is reported to be 54 ps[34], not far from the estimated time of sensitization by bright exciton in Mn-doped $CsPbBr_3$ NPLs in our study (~50 ps). Therefore, the sensitization should occur from both bright and dark exciton at 10 K, since the thermal excitation from dark to bright state is essentially prohibited. In contrast, the sensitization at 170 K should be largely via bright exciton due to the facile thermal equilibration between bright and dark state. For this comparison, we use the kinetic scheme shown in Fig. 5e and examine $I_{Mn,\tau}/(I_{ex}/I_{ex}^0)$ for two different cases, i.e., one involving bright exciton (170 K) and the other involving both bright and dark excitons (10 K), which are detailed in Supplementary information. The comparison indicates that the similar $I_{Mn,\tau}/(I_{ex}/I_{ex}^0)$ at 10 K and 170 K requires similar values of $k_{ET,B}/k_{ex,r,B}$ and $k_{ET,D}/k_{ex,r,D}$, i.e., the comparable sensitizing capabilities of bright and dark exciton while the value of each rate constant between bright and dark exciton may differ by many orders of magnitude. Although the possible structural heterogeneity of the dopant and defect sites (e.g., interior vs surface) are not well understood and adds the kinetic heterogeneity not accounted for in the employed kinetic model, the above analysis is sufficient to conclude the active role of dark exciton in sensitization of $Mn^{2+}$.

Unlike the sensitization rate, the Auger energy transfer rate is more difficult to estimate from the PL measurements. However, the relative change in Auger energy transfer rate with varying temperature can be inferred from the temperature-dependent $J_{hot}$ and PL data. In Fig. 3f, it is notable that $J_{hot}$ of Mn-doped CsPbBr$_3$ NPLs increases continuously with cooling below 120 K unlike the nearly constant $I_{Mn,\tau}/(I_{ex}/I_{ex}^0)$ in Fig. 4d. Considering that the total PL intensity, $\tau_{Mn}$ and $I_{Mn,\tau}/(I_{Mn}/I_{ex}^0)$ are nearly constant below 120 K, the continuing increase of $J_{hot}$ cannot be due to the change in the sensitization nor nonradiative decay of exciton and $|Mn^{2+}\rangle^*$. One can also rule out the potential increase of hot electron energy with cooling as the possible origin, since both exciton and Mn$^{2+}$ PL show a small redshift (< 10 meV) with cooling in this temperature range that should have the opposite effect if there is any. (Fig. S6). The increasing $J_{hot}$ below 120 K can then be interpreted as the more effective Auger energy transfer upon cooling. Since the Auger energy transfer should involve both bright and dark excitons at low temperatures as in sensitization, it indicates that the dark exciton is active in the Auger energy transfer and may even be more effective than bright exciton. Clearly, better understanding of how the longer lifetime and the weaker transition dipole of dark exciton affect the sensitization and Auger energy transfer in future studies will be crucial in exploring the potential benefits of dark exciton-mediated upconversion.

Compared to Mn-doped CsPbBr$_3$ NPLs, Mn-doped CsPbBr$_3$ QDs show a more complex temperature dependence of $I_{Mn,\tau}/(I_{ex}/I_{ex}^0)$ and $J_{hot}$. The analysis of the temperature-dependent $I_{Mn,\tau}/(I_{ex}/I_{ex}^0)$ of Mn-doped CsPbBr$_3$ QDs (Fig. S9) suggests more efficient sensitization by dark exciton than bright exciton, although the firmer conclusion requires further study. $J_{hot}$ shows a dip near 200 K that is absent in the temperature-dependent sensitization, which is reminiscent of the thermally activated trapping of exciton competing with radiative recombination of exciton or sensitization in the studied earlier in other QDs.[24, 32, 37, 38] The distinct dip in $J_{hot}$ in Mn-doped CsPbBr$_3$ QDs suggests that Auger energy transfer in the QD host may be more susceptible to the charge carrier trapping than the sensitization is. Interestingly, Mn-doped CsPbCl$_3$ QDs also exhibit a dip in the temperature-dependent $J_{hot}$ (Fig. S7b), indicating the importance of the host morphology in determining hot electron upconversion behavior. We conjecture that the Auger energy transfer in NPLs may have less influence from the competing thermally activated charge carrier trapping than in the QDs due to the stronger exciton-dopant interaction, which is another merit of 2-dimensional host nanocrystals over the QDs. In this regard, a systematic study in Mn-doped NPLs of varying thicknesses including the monolayer structure will provide further insights into the advantages of 2-dimensional MHP nanocrystals in generating hot electrons via upconversion.

In conclusion, we demonstrated the photoemission of the upconverted hot electrons in Mn-doped CsPbBr$_3$ QDs and NPLs beyond Mn-doped II-VI QDs indicating that hot electron upconversion can be realized more universally in various host semiconductor nanocrystals. In particular, few monolayer-thick Mn-doped $CsPbBr_3$ NPLs showed the advantages over QDs due to the stronger exciton-dopant interaction resulting in the faster saturation of the intermediate state and less susceptibility to the charge carrier trapping in hot electron upconversion. Furthermore, Mn-doped $CsPbBr_3$ NPLs showed evidence for the hot electron upconversion from the long-lived dark exciton prevailing, revealing a new pathway deserving further exploration that may potentially enhance the upconversion efficiency. The insights obtained from this study will be valuable for obtaining the energetic upconverted hot electrons more efficiently through the rational design of the doped semiconductor nanostructures.

**Methods**

*Synthesis of Mn-doped $CsPbBr_3$ nanocrystals*

Mn-doped $CsPbBr_3$ QDs (6 nm, ~10 % doping) were prepared using previously reported method with modification.[22] In short, Cs-oleate was injected into Pb-Mn precursor heated at 200 °C to initiate the formation of nanocrystals. Afterward, acetone was used to precipitate the QDs and hexane was used to resuspend the QDs after centrifugation. Mn-doped $CsPbBr_3$ NPLs (~2 nm thick, ~10 % doping) were prepared via halide exchange of Mn-doped $CsPbCl_3$ NPLs serving as the template, which were synthesized using the previously reported method.[39] Bromotrimethylsilane (TMSBr) was used as the source of bromide when performing the halide exchange reaction.[40] Further details on the synthesis and characterization of the nanocrystals size and morphology are in the Supplementary Information.

*Measurements of steady state optical spectra and time-resolved photoluminescence*

Steady state absorption and photoluminescence spectra of Mn-doped $CsPbBr_3$ nanocrystals under varying were obtained both in solution phase and as a polystyrene-dispersed nanocrystal film on a sapphire substrate. For the temperature-dependent PL measurement, the nanocrystal film on the substrate was mounted in an optical cryostat that allow the temperature control within 10-300 K. A CCD spectrometer was used to obtain all steady state optical spectra. For the time-resolved exciton PL measurement under varying temperature, the PL signal from the film samples mounted in the cryostat was recorded via time-correlated single photon counting. The time-resolved exciton PL under the external magnetic field was obtained using a magneto-optical cryostat (CryoVac) in conjunction with a superconducting magnet that can provide the external magnetic field of 0-8 T. Further details of the measurement and sample preparation are in the supplementary information.

*Hot electron photoemission current measurement*

Photoemission current of hot electrons was measured in an optical cryostat (ST-100, JANIS) that provides the temperature control in 10-300 K. The substrates were prepared following the procedure stated in supplementary information and mounted onto the cold finger in the configuration illustrated in Fig. 1b. A 405 nm *cw* laser was focused onto the substrate and the generated photoelectron was collected by a copper counter electrode as shown in Fig. 1b. Photocurrent from the hot electron was detected with a picoammeter (Model 6487, Keithley) in conjunction with a lock-in amplifier. For the bias-dependent photocurrent measurement, a DC voltage generated by the picoammeter was applied between copper and ITO electrode. Further details of the measurement are in supplementary information.

**Acknowledgement**

This work is supported by National Science Foundation (CHE-200396). We thank Doyun Kim and Prof. Qing Tu for the assistance in AFM measurement.

**Author Information**

Affiliation

**Department of Chemistry, Texas A&M University, College Station, TX, USA**

Chih-Wei Wang, Tian Qiao, Dong Hee Son

**Department of Physics, Texas A&M University, College Station, TX, USA**

Xiaohan Liu, Mohit Khurana, Alexey V. Akimov

**Center for Nanomedicine, Institute for Basic Science and Graduate Program of Nano Biomedical Engineering, Advanced Science Institute, Yonsei University, Seoul, Republic of Korea**

Dong Hee Son

Contribution

C.-W.W. and T.Q. carried out material syntheses, experiment and data analysis. X.L. and M.K. carried out magneto PL measurements and analysis. A.V.A. and D.H.S. conceived and desinged experiment. C.-W.W., A.V.A. and D.H.S. wrote the manuscript.

Corresponding authors

Correspondence to Dong Hee Son

**Ethics declarations**

Competing interests

The authors declare no competing interests.

**Supplementary Information**

Supplementary Figs. S1-S9, Scheme S1-S4, Table S1-S2, Detailed synthesis and measurement procedure, kinetic modeling.

Supplementary Information for

# Photoemission of the Upconverted Hot Electrons in Mn-doped CsPbBr$_3$ Nanocrystals

Chih-Wei Wang[1], Xiaohan Liu[2], Tian Qiao[1], Mohit Khurana[2], Alexey V. Akimov[2], Dong Hee Son[1,3]

[1]Department of Chemistry, Texas A&M University, College Station, TX, USA
[2]Department of Physics, Texas A&M University, College Station, TX, USA
[3]Center for Nanomedicine, Institute for Basic Science and Graduate Program of Nano Biomedical Engineering, Advanced Science Institute, Yonsei University, Seoul, Republic of Korea

**Table of Contents**

1. **Materials synthesis**

   a. **Chemicals**

   Cesium carbonate (Alfa Aesar), lead (II) chloride (Alfa Aesar), lead (II) bromide (Alfa Aesar), manganese (II) chloride tetrahydrate (Sigma Aldrich), manganese (II) acetate (Sigma Aldrich), hydrobromic acid (Alfa Aesar), bromotrimethylsilane (Sigma Aldrich), 1-octadecene (Sigma Aldrich), oleylamine (Sigma Aldrich) and oleic acid (Sigma Aldrich) were purchased from suppliers and used without further purification.

   b. **Synthesis of 6 nm Mn-doped $CsPbBr_3$ quantum dots (QDs)**

   Mn-doped $CsPbBr_3$ quantum dots (6 nm) were synthesized by following the previously reported method with some modifications.[S1] Initially, a mixture of $PbBr_2$ (60 mg), $Mn(OAc)_2$ (180 mg), HBr (500 μL), oleic acid (0.7 mL), oleylamine (0.7 mL) and 1-octadecene (5 mL) was degassed under vacuum at 140 °C for 1 hour in a 50 mL 3-neck flask, and the flasks was purged with nitrogen. Subsequently, 0.7 mL of oleic acid and oleylamine were injected to this mixture and heated at 200 °C for 1 hour. Maintaining the temperature of the mixture in the flask at 200 °C, 0.4 mL of Cs-oleate solution was swiftly injected into the mixture, which initiated the formation of the quantum dots. The reaction was quenched after 5 seconds of reaction by cooling the flask in an ice bath. The synthesized QDs were recovered by precipitating with acetone and resuspended in hexane. To purify the QDs, methyl acetate was added into the solution to precipitate the QDs. The QDs are isolated via centrifugation and redispersed in hexane. In this synthesis, Cs-oleate solution was prepared by heating the mixture of cesium carbonate (250 mg), oleic acid (800 mg) and 1-octadecene (7 g) under vacuum at 120 °C for 30 minutes. During the synthesis, Cs-oleate solution was kept under nitrogen at 130 °C to avoid solidification.

   c. **Synthesis of 2 nm-thick Mn-doped $CsPbBr_3$ nanoplatelets (NPLs)**

   Mn-doped $CsPbBr_3$ NPLs were synthesized in two steps: (1) synthesis of Mn-doped $CsPbCl_3$ NPLs [S2], (2) halide exchange to convert Mn-doped $CsPbCl_3$ NPLs to Mn-doped $CsPbCl_3$ NPLs.[S3]. For the synthesis of Mn-doped $CsPbCl_3$ NPLs, initially a mixture of $PbCl_2$ (60 mg), $MnCl_2·4H_2O$ (180 mg), $CuCl_2$ (240 mg), oleic acid (2 mL), oleylamine (2 mL) and 1-octadecene (5 mL) was degassed under vacuum at 120 °C for 30 minutes in a 50 mL 3-neck flask. After purging with nitrogen, the mixture was heated to 200 °C and held for 5 minutes. The flask was then allowed to cool down in a water bath. During cooling, 0.6 mL of the aforementioned Cs-oleate solution was added into the mixture. To initiate the formation of NPLs, 10 mL of acetone was added into the mixture rapidly. The NPLs were separated from the solution via centrifugation followed by resuspension in hexane. To purify the NPLs, acetone was added into the solution to precipitate the nanocrystals. The NPLs were isolated via centrifugation and resuspended in hexane. The resulting Mn-doped $CsPbCl_3$ NPLs is ~2 nm in thickness corresponding to 3 monolayer, and 30-60 nm in lateral dimension. Using Mn-doped $CsPbCl_3$ NPLs as the precursor nanocrystals, Mn-doped $CsPbBr_3$ NPLs were prepared via post-synthesis halide exchange. To increase the stability of the purified $CsPbCl_3$ NPLs before the halide exchange process, 200 μL of toluene solution of didodecyldimethylammonium chloride (DDAC) was added into the NPL suspension. Excess DDAC was removed by the precipitation of NPLs through addition of methyl acetate. The precipitate was isolated via centrifugation and resuspended in hexane. Halide exchange was carried out by adding 100 μL of neat bromotrimethylsilane (TMSBr) into the Mn-doped $CsPbCl_3$ NPLs solution. The mixture was stirred for 15 minutes. After the completion of the reaction, the excess TMSBr was removed *in vacuo*. The converted NPLs were dispersed in heptane for further use.

d. **Synthesis of Mn-doped $CsPbCl_3$ QDs**

Mn-doped $CsPbCl_3$ QDs were synthesized following the previously reported procedure.[S4] A mixture of $PbCl_2$ (28 mg), $MnCl_2·4H_2O$ (10 mg), oleic acid (0.5 mL), oleylamine (0.5 mL) and 1-octadecene (5 mL) was degassed under vacuum at 120 °C for 30 minutes in a 50 mL 3-neck flask before purging with nitrogen. Trioctylphosphine (0.2 mL) and oleylammonium chloride (0.5 mL, 10% in oleylamine) was injected and the temperature was raised to 180 °C for 30 minutes. 0.4 mL of Cs-oleate solution was swiftly injected into the mixture and the reaction was quenched within 5 seconds using an ice bath. The QDs was then precipitated with acetone for purification. The precipitate was isolated via centrifugation and resuspended in hexane for further use.

Table S1. PL properties of Mn-doped $CsPbBr_3$ nanocrystals in solution at the ambient temperature.

| Sample | $\lambda_{ex}$ (nm) | $\lambda_{Mn}$ (nm) | $\tau_{Mn}$ (μs) | $QY_{ex}$ (%) | $QY_{Mn}$ (%) | $Mn^{2+}$ Conc. (%) |
|---|---|---|---|---|---|---|
| NPLs | 440 | 613 | 411 | ~0.5 | 32.5±2 | ~10 |
| QDs | 490 | 616 | 261 | 8±1 | 32±2 | ~10 |

$\lambda_{ex}$: exciton PL wavelength, $\lambda_{Mn}$: $Mn^{2+}$ PL wavelength, $\tau_{Mn}$: $Mn^{2+}$ PL lifetime,
$QY_{ex}$: exciton PL quantum yield, $QY_{ex}$: $Mn^{2+}$ PL quantum yield

## 2. Sample Characterization

### a. Transmission Electron Microscopy of Nanoparticles

Images of the nanocrystals used in this study was captured with FEI Tecnai G2 F20 ST field-emission microscope. The sample was prepared by drying the nanocrystal solution onto the TEM grid *in vacuo*. Lateral size of the NPLs was estimated from the length distribution from the TEM images.

### b. Elemental Analysis

To determine the doping concentration and estimate the absorption cross section used in the kinetic modeling, inductively coupled plasma mass spectrometry (Nexlon 300D) was employed to obtain the information of the elemental composition of the particles. Nanocrystals were digested in 70% $HNO_{3(aq)}$ and diluted to appropriate concentrations. Standard solutions of $Cs^+$, $Pb^{2+}$, $Mn^{2+}$ (Sigma Aldrich, 1000 mg/L) were used to construct the calibration curves to determine the elemental composition.

### c. Atomic force microscopy of nanocrystal film

The resulting deposited QDs and NPLs were characterized by tapping mode AFM with a diamond-like-carbon coated cantilever (AIODLC, nominal spring constant 2.7 N/m) using an MFP-3D Infinity AFM instrument (Asylum Research).

## 3. Optical measurements

### a. Measurement of absorption and photoluminescence (PL) spectra

Absorption spectra of the nanocrystal solution at the ambient temperature were collected with a CCD spectrometer (USB2000, Ocean Optics) using heptane as the solvent. Solution-phase PL spectra at the ambient temperature were collected with a CCD spectrometer (QE65, Ocean Optics) under 380 nm excitation from a xenon lamp (Model 67005, Oriel Instrument) in conjunction with a monochromator (1/8 m Monochromator, Oriel Cornerstone).

For the temperature-dependent PL measurements, polymer-dispersed nanocrystals deposited on sapphire substrates were used. The nanocrystals were initially dispersed in 2% (w/w) polystyrene toluene solution and deposited on sapphire substrate by drop casting. The deposited nanocrystal samples on sapphire substrate were dried under vacuum for over 30 minutes before placing in an optical cryostat (JANIS/CryoVac). For the spectra shown in Fig. S3, from which the data in Fig. 4 is extracted, a temperature controller (Model 335, LakeShore) was used to control the temperature in 10-300K range. A 405 nm *cw* diode laser was used to excite the nanocrystal samples and the PL was recorded with a CCD spectrometer (QE65, Ocean Optics) coupled to an optical fiber and an achromatic lens. For the higher resolution spectra of Mn-doped $CsPbBr_3$ NPLs shown in Fig. 4a, spectrometer (PIXIS 100F, Princeton Instrument) with much higher resolution (1200 gr/mm$^2$) was used for recording the spectra that enabled spectrally resolving the bright and dark exciton PL.

b. **Time-resolved measurements of PL and magneto-PL**

Several different time-resolved measurements of PL were performed using different setups and detectors. For all time-resolved PL measurements, the nanocrystal samples were prepared on the sapphire substrate in polystyrene matrix as described in the section above.

For the time-resolved measurement of $Mn^{2+}$ PL, from which the lifetime of excited ligand field state of $Mn^{2+}$ ($|Mn^{2+}\rangle^*$) were obtained using a 337 nm pulsed nitrogen laser (NL100, Stanford Research Systems) as excitation source. The time-dependent PL signal detected with a photomultiplier tube (R928, Hamamatsu Photonics KK) and an amplifier (C9663, Hamamatsu Photonics KK) was recorded using a digital oscilloscope (WaveAce 234, Teledyne Lecroy). For the temperature-dependent measurement, identical optical cryostat/temperature controller setup in section 3.a was used for the temperature control within 10-300 K range.

Time-resolved exciton PL of Mn-doped $CsPbBr_3$ NPLs under varying temperatures and under varying magnetic fields were measured using a magneto-optical cryostat (Cryovac). The magnetic field was generated by superconducting magnets up to 8 T. For the magneto fluorescence measurements examining the magnetic field-dependent dark exciton lifetime at 10 K, the fluorescence signal was detected in Faraday geometry. A pulsed 405 nm laser with pulse width of 150 ns was used for the excitation. The samples were excited at the fluence of ~30 μW/cm$^2$ to ensures the excitation density of much less than 1 exciton/nanocrystal during the lifetime of ($|Mn^{2+}\rangle^*$). Time-resolved PL intensity was recorded via time-correlated single photon counting (PicoHarp 300) with an avalanche photodiode (MPD PDM series).

## 4. Fabrication of electrodes and photoemission current measurement

Strips (1 cm × 4 cm) of ITO/glass substrates (Thin Film Device) were used as the substrate for the deposition of the nanocrystals. Before depositing the nanocrystals, ITO/glass substrates were cleaned by soaking in piranha solution and sonicating in isopropanol for 20 minutes. The cleaned substrates were dried in an oven before use. The photocathode was fabricated by spin coating the nanocrystals using a concentrated nanocrystal solution dispersed in heptane. The nanocrystal film on ITO/glass substrate had relatively uniform surface topography as examined by atomic force microscopy as shown in Fig. S8. A copper plate of the same size as the ITO/glass substrate was used as the electrode collecting the hot electrons photoemitted from the photocathode. The photocathode and copper electrode were assembled with 2 mm spacing between the two electrodes using Kapton tape as the insulating spacer. For the electrical connection to each electrode and the picoammeter, a copper wire was connected to each electrode.

For the measurement of photoemission current, the photocathode side of the assembled electrodes was directly mounted on the cold finger of an optical cryostat (ST-100, JANIS) for the measurements under varying temperature, excitation intensity and electrical bias between the two electrodes. A 405 nm *cw* diode laser (FBB-405, RGBLase LLC.) was used as the excitation source. To measure the beam size, a CCD camera was used to capture the 2-dimensional intensity profile of the beam. The beam size at the sample substrate is ~200 μm in diameter at full width at half maximum intensity, which was used to obtain the photoemission current density from the measured photoemission current. The hot electrons photocurrent was measured using a picoammeter (Model 6487, Keithley) and a lock-in amplifier (SR830, SRS) in conjunction of an optical chopper that modulates the excitation light at 140 Hz. To minimize the environmental noise, a Faraday cage was placed to shield the electrical noise. The reported photoemission current density at a given experimental condition is the average of 5 different independent measurements at 5 different locations on the electrode to avoid photodegradation of the nanocrystals.

## 5. Relationship between $I_{Mn,\tau}/(I_{ex}/I_{ex}^0)$ and $k_{ET}/k_{ex,r}$

In a simple kinetic model shown, below where the exciton ($|ex\rangle$) decays via radiative and nonradiative recombination and sensitization that populates the excited $Mn^{2+}$ ligand field state ($|Mn^{2+}\rangle^*$), the intensities of $Mn^{2+}$ PL normalized to $Mn^{2+}$ PL lifetime ($I_{Mn,\tau}$) and exciton PL ($I_{ex}$) are expressed as follows. The rate constants in scheme S1 are defined as follows.

- $k_{ex,r}$ : radiative recombination of $|ex\rangle$
- $k_{ex,nr}$ : nonradiative recombination of $|ex\rangle$
- $k_{ET}$ : sensitization
- $k_{Mn,r}$ : radiative relaxation of $|Mn^{2+}\rangle^*$
- $k_{Mn,nr}$ : nonradiative relaxation of $|Mn^{2+}\rangle^*$

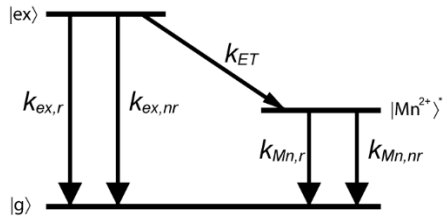

Scheme S1. Three states kinetic model.

$Mn^{2+}$ PL intensity ($I_{Mn}$) can be expressed as follows using the isolated quantum yield of $|Mn^{2+}\rangle^*$ ($QY_{Mn}$) defined as below.

$$I_{Mn} \propto \frac{k_{ET}}{k_{ex,r} + k_{ex,nr} + k_{ET}} \times QY_{Mn}$$

$$QY_{Mn} = \frac{k_{Mn,r}}{k_{Mn,r} + k_{Mn,nr}}$$

(Eq. 1)

$I_{Mn,\tau}$ can be expressed as follows from the lifetime of $|Mn^{2+}\rangle^*$, $\tau_{Mn} = 1/(k_{Mn,r} + k_{Mn,nr})$. $a$ is $\tau_{Mn}/I_{ex}^0$ with the value of $\tau_{Mn}$ at 10 K following its definition in the manuscript.

$$I_{Mn,\tau} = a \frac{I_{Mn}}{\tau_{Mn}} = a \frac{k_{ET}}{k_{ex,r} + k_{ex,nr} + k_{ET}} \times \frac{k_{Mn,r}}{k_{Mn,r} + k_{Mn,nr}} \times (k_{Mn,r} + k_{Mn,nr})$$

$$= a \frac{k_{ET} k_{Mn,r}}{k_{ex,r} + k_{ex,nr} + k_{ET}}$$

(Eq. 2)

$I_{ex}$ is expressed as follows.

$$I_{ex} \propto \frac{k_{ex,r}}{k_{ex,r} + k_{ex,nr} + k_{ET}} \quad (Eq.\ 3)$$

If we assume that $QY_{Mn}$ is close to 1 at sufficiently low temperature, e.g., 10 K, $\tau_{Mn}=1/k_{Mn,r}$ and $a = 1/(\tau_{Mn} \times I_{ex}^0)$. Under this assumption, we get the relationship $I_{Mn,\tau}/(I_{ex}/I_{ex}^0) = k_{ET}/k_{ex,r}$ as follows.

$$\frac{I_{Mn,\tau}}{I_{ex}/I_{ex}^0} = \frac{a \dfrac{k_{ET}k_{Mn,r}}{k_{ex,r} + k_{ex,nr} + k_{ET}}}{(\dfrac{k_{ex,r}}{k_{ex,r} + k_{ex,nr} + k_{ET}})/I_{ex}^0} = \frac{k_{ET}}{k_{ex,r}} \quad (Eq.\ 4)$$

## 6. Comparison of $I_{Mn,\tau}/(I_{ex}/I_{ex}^0)$ at two different temperatures

For the comparative analysis of $I_{Mn,\tau}/(I_{ex}/I_{ex}^0)$ at 10 K and 170 K, we use the kinetic scheme shown in Fig. 5e. We also make the following assumptions for the comparison. (1) At 170 K, we can ignore the involvement of dark exciton in sensitization, (2) Only at 10 K, both bright and dark exciton are involved in sensitization. Since the absolute values of rate constant can vary with temperature and currently unknown, we will keep the ratio constant regardless of the temperature. Because of the uncertainly of the rate constants and actual quantum yield of the PL, the analysis is not intended to be quantitative but aimed to show that $k_{ET,B}/k_{ex,r,B}$ and $k_{ET,D}/k_{ex,r,D}$ should be similar to for $I_{Mn,\tau}/(I_{ex}/I_{ex}^0)$ to be similar at 10 and 170 K, despite the limitation of the model and assumptions made to simplify the problem.

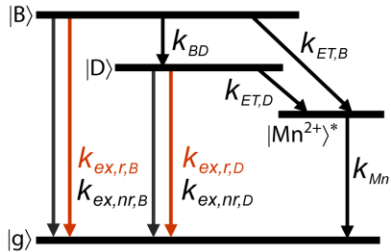

Scheme S2. Four states kinetic model.

**Case 1 (170 K)**

$I_{Mn,\tau}/(I_{ex}/I_{ex}^0)$ can be expressed by using simpler scheme ignoring the involvement of dark exciton following the derivation in page S7-S8.

$$\frac{I_{Mn,\tau}}{I_{ex}/I_{ex}^0} = \frac{k_{ET,B}}{k_{ex,r,B}} \qquad (Eq.\ 5)$$

For Mn-doped NPLs, we have $k_{ET,B}: k_{ex,r,B} = 20:1$, since $I_{Mn,\tau}/(I_{ex}/I_{ex}^0)$ =20 from Fig. 5d.

**Case 2 (10 K)**

Mn$^{2+}$ PL intensity ($I_{Mn}$) has two contributions, one from sensitization by bright exciton and the other from sensitization by dark exciton. Therefore, $I_{Mn,\tau}$ can be expressed as follows by using the same procedure as in Eq. 4, where $a = \frac{1}{\tau_{Mn} \times I_{ex}^0}$.

$$I_{Mn,\tau} \propto \left( \frac{k_{ET,B}}{k_{ex,r,B} + k_{ex,nr,B} + k_{BD} + k_{ET,B}} + \frac{k_{BD}\left(\frac{k_{ET,D}}{k_{ex,r,D} + k_{ex,nr,D} + k_{ET,D}}\right)}{k_{ex,r,B} + k_{ex,nr,B} + k_{BD} + k_{ET,B}} \right) \frac{k_{Mn,r}}{\tau_{Mn} \times I_{ex}^0} \qquad (Eq.\ 6)$$

$I_{ex}/I_{ex}^0$ also has two contributions.

$$\frac{I_{ex}}{I_{ex}^0} \propto \left( \frac{k_{ex,r,B}}{k_{ex,r,B} + k_{ex,nr,B} + k_{BD} + k_{ET,B}} + \frac{k_{BD}\left(\frac{k_{ex,r,D}}{k_{ex,r,D} + k_{ex,nr,D} + k_{ET,D}}\right)}{k_{ex,r,B} + k_{ex,nr,B} + k_{BD} + k_{ET,B}} \right) / I_{ex}^0 \qquad (Eq.\ 7)$$

Therefore, $I_{Mn,\tau}/(I_{ex}/I_{ex}^0)$ is expressed as follows.

$$\frac{I_{Mn,\tau}}{I_{ex}/I_{ex}^0} = \frac{k_{ET,B} + k_{BD}\left(\frac{k_{ET,D}}{k_{ex,r,D} + k_{ex,nr,D} + k_{ET,D}}\right)}{k_{ex,r,B} + k_{BD}\left(\frac{k_{ex,r,D}}{k_{ex,r,D} + k_{ex,nr,D} + k_{ET,D}}\right)} \left(\frac{k_{Mn,r}}{\tau_{Mn}}\right) \qquad (Eq.\ 8)$$

For Mn-doped NPLs, we make the same assumption again $k_{ET,B}: k_{ex,r,B} = 20:1$. In addition, we will also assume that $k_{ET,B}$ and $k_{BD}$ are similar as discussed in the manuscript, with the overall assumption of $k_{ET,B}: k_{BD}: k_{ex,r,B}=20:20:1$. If the nonradiative decay channels can be ignored at 10 K, Eq. 8 becomes

$$\frac{I_{Mn,\tau}}{I_{ex}/I_{ex}^0} = \frac{20 + 20\left(\frac{k_{ET,D}}{k_{ex,r,D} + k_{ET,D}}\right)}{1 + 20\left(\frac{k_{ex,r,D}}{k_{ex,r,D} + k_{ET,D}}\right)} \quad \text{(Eq. 9)}$$

For the value of $I_{Mn,\tau}/(I_{ex}/I_{ex}^0)$ in Eq. 9 to be 20 at 10 K (from Fig. 5d), solving the equation gives $k_{ET,D}: k_{ex,r,D} = 20:1$, same as in the case of bright exciton in Eq. 5.

### 7. Sensitization rate in Mn-doped CsPbBr$_3$ QDs and NPLs

There are several studies reporting the sensitization rates in various Mn-doped perovskite nanocrystals. However, the reported rates vary significantly depending on the host composition, doping concentration, size and morphology of the host nanocrystal, and the method of determining the rate. In 10 nm 0.4 % Mn-doped CsPbCl$_3$ QDs, sensitization time of 380 ps was reported from the transient absorption measurement.[S5] In 2-monolayer thick, 2 % Mn-doped MAPbCl$_3$ NPLs, sensitization time of 60 ps was obtained from the analysis of PL data.[S6] For Mn-doped CsPbBr$_3$ NPLs (doping concentration unknown), one study reported less than 1 ps time for sensitization.[S7] In the present study, the approximate sensitization rate constants ($k_{ET}$) were obtained from the measured exciton PL lifetime and exciton PL quantum yield in undoped nanocrystals in conjunction with the relationship $I_{Mn,\tau}/(I_{ex}/I_{ex}^0) = k_{ET}/k_{ex,r}$ in Mn-doped nanocrystals discussed in Section 5 of the Supplementary Information. The radiative exciton recombination rate constants ($k_{ex,r}$) of undoped CsPbBr$_3$ QDs (6 nm) and NPLs (2 nm thick) at room temperature are 1/5.4 ns$^{-1}$ and 1/3.3 ns$^{-1}$ respectively for the QDs and NPLs based on the measured PL lifetime and quantum yield.[S8, S9] The radiative rates are assumed to be the same in Mn-doped QDs and NPLs. Values of $I_{Mn,\tau}/(I_{ex}/I_{ex}^0)$ for Mn-doped CsPbBr$_3$ QDs and NPLs are 5-10 and 70 respectively. Therefore, the estimated sensitization rate is 1/1080 - 1/540 ps$^{-1}$ and 1/47 ps$^{-1}$ for QDs and NPLs respectively that differ by an order of magnitude.

### 8. Modeling of steady state population of $|Mn^{2+}\rangle^*$ in Mn-doped CsPbBr$_3$ QDs and NPLs.

To examine the different saturation behavior of $|Mn^{2+}\rangle^*$ state in Mn-doped CsPbBr$_3$ QDs and NPLs, intensity-dependent steady state population of $|Mn^{2+}\rangle^*$ was calculated using the following simple kinetic model and approximate rate constants. The sensitization of $Mn^{2+}$ is unidirectional in this model as the difference in exciton and Mn transition energies (>0.5 eV) is much larger than thermal energy (kT) in our experimental conditions. Because of the uncertainties on the rate constants, the comparison is intended to verify that Mn-doped CsPbBr$_3$ NPLs can reach the saturation of $|Mn^{2+}\rangle^*$ population in the employed excitation intensity range in contrast to the QDs rather than providing quantitatively accurate simulation.

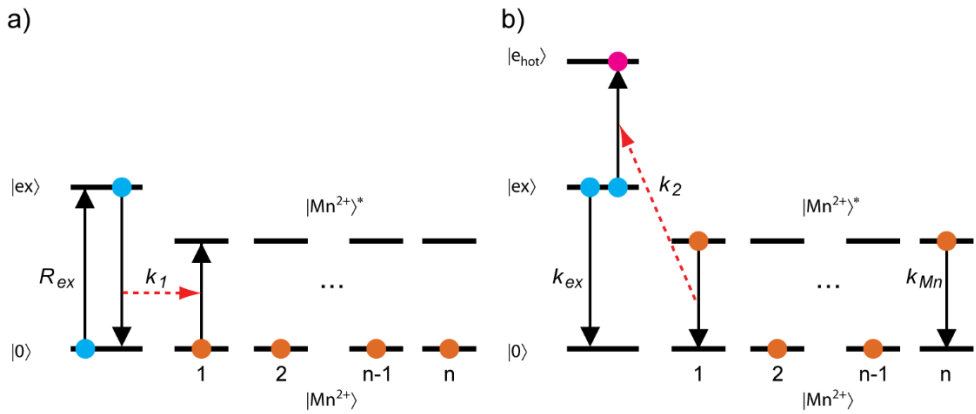

Scheme S3. Kinetic processes involved in the modeling. (a) Population processes and (b) depopulation processes.

The kinetic parameters are defined as follows.

$R_{ex}$ : photoexcitation rate (s$^{-1}$) = photon flux × absorption cross section (σ)

$k_1$ : sensitization rate between exciton and one $|Mn^{2+}\rangle$ = $k_{ET}/N_{Mn}^0$

$k_2$ : Auger energy transfer rate between one $|Mn^{2+}\rangle^*$ and $|ex\rangle$ = $k_{Aug}/N_{Mn}^0$

$k_{ex}$ : exciton recombination rate

$k_{Mn}$ : relaxation rate of $|Mn^{2+}\rangle^*$

$N_{ex}$ : population of $|ex\rangle$

$N_{Mn}^0$ : number of $Mn^{2+}$ dopants

$N_{Mn}$ : population of $|Mn^{2+}\rangle$

$N_{Mn}^0 - N_{Mn}$: population of $|Mn^{2+}\rangle^*$

The steady state approximation of exciton and $Mn^{2+}$ population gives the following set of equations, which can be solved to give the steady state population of $|Mn^{2+}\rangle^*$ as a function of $R_{ex}$.

$$\frac{dN_{ex}}{dt} = R_{ex} - k_{ex}N_{ex} - k_1 N_{ex} N_{Mn} - k_2 N_{ex}(N_{Mn}^0 - N_{Mn}) = 0$$

$$\frac{d(N_{Mn}^0 - N_{Mn})}{dt} = k_1 N_{ex} N_{Mn} - k_{Mn}(N_{Mn}^0 - N_{Mn}) - k_2 N_{ex}(N_{Mn}^0 - N_{Mn}) = \quad \text{(Eq. 10)}$$

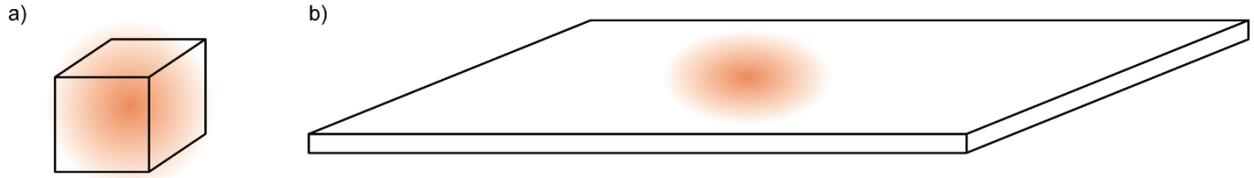

Scheme S4. Illustration of volume encompassed by the exciton with 3.5 nm Bohr radius in (a) QDs (6 nm) and (b) NPLs (2 nm×25 nm×45 nm).

Table S2 summarizes the parameters used the calculation. Volume (V) for QD is the volume of 6 nm QD. V for NPLs is the volume encompassed by 2-D exciton with Bohr radius ($a_B$) of 3.5 nm as illustrated above (V= 2 × $\pi a_B^2$ (nm$^3$) = 77 nm$^3$) not the total volume, assuming that only those $Mn^{2+}$ ions within the volume encompassed by one exciton are relevant in the sensitization and Auger energy transfer in NPLs. $N_{Mn}^0$, σ and $R_{ex}$ for NPLs in Table S2 were also obtained using this V. σ of QDs that determine $R_{ex}$ was taken from a literature.[S10] For Mn-doped CsPbBr$_3$ NPLs, elemental analysis approach described in Ref. S10 was employed using the average size of the Mn-doped CsPbBr$_3$ NPLs (2 nm×25 nm×45 nm) determined from TEM images. The volume normalized σ of Mn-doped CsPbBr$_3$ NPLs at 400 nm determined in this way is ~1.5 times larger than that of Mn-doped CsPbBr$_3$ QDs. Since $k_2$ is unknown, the range of the values spanning one order of magnitude (0.1-1$k_1$) were used. $k_{Mn}$ was measured experimentally. $N_{Mn}^0$ is for the nanocrystals of volume V in Table S2 and 10 % B-site substitutional doping concentration.

Table S2. Parameters used for the intensity-dependent steady state $|Mn^{2+}\rangle^*$ population.

|  | V (nm³) | σ (cm²) at 405 nm | $R_{ex}$ (s⁻¹) for 1 W/cm² | $k_{ET}$ (s⁻¹) | $k_1$ (s⁻¹) | $k_2$ (s⁻¹) | $k_{Mn}$ (s⁻¹) | $N_{Mn}^0$ |
|---|---|---|---|---|---|---|---|---|
| QDs | 216 | 2.0 ×10⁻¹⁴ | 4.1 ×10⁴ | 1.0 × 10⁹ | 1.0 × 10⁷ | 0.1 -1 $k_1$ | 4.0 ×10³ | 100 |
| NPLs | 77 | 1.1 ×10⁻¹⁴ | 2.2 ×10⁴ | 2.0 × 10¹⁰ | 5.5 × 10⁸ | 0.1 -1 $k_1$ | 2.5 ×10³ | 36 |

Using the solution of the coupled rate equations in Eq. 10 and the parameters in Table S2, we compared the intensity-dependent steady state population of $|Mn^{2+}\rangle^*$ in Mn-doped CsPbBr₃ QDs and NPLs as shown in Fig. S2. The comparison indicates that NPLs exhibit much faster saturation of $|Mn^{2+}\rangle^*$ population than QDs.

**Additional Figures**

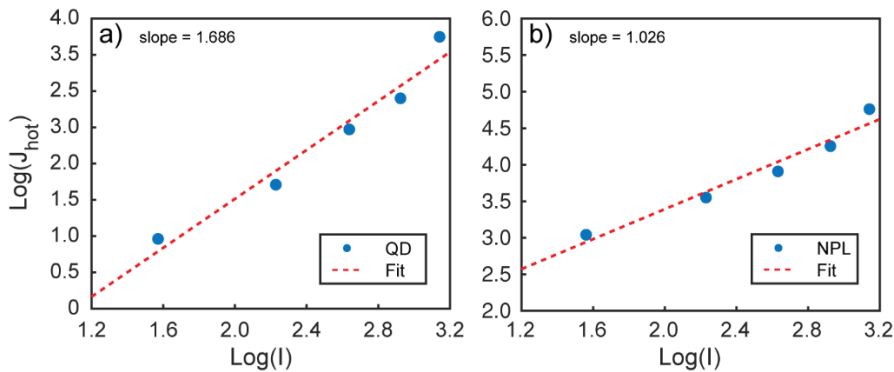

Figure S1. log-log plot of the intensity dependent $J_{hot}$ of (a) CsPbBr$_3$ QDs and (b) NPLs shown in Fig.3a and b. The dashed lines are the linear fit of the plots.

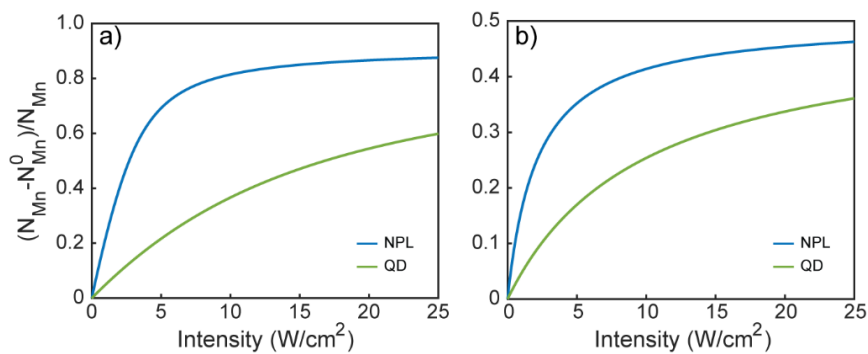

Figure S2. Steady state population of excited Mn$^{2+}$ state for Mn-doped CsPbBr$_3$ under cw excitation with two Auger rate. (a) $k_2 = 0.1k_1$ (b) $k_2 = k_1$. The details of the calculation are in section 8 of Supplementary information.

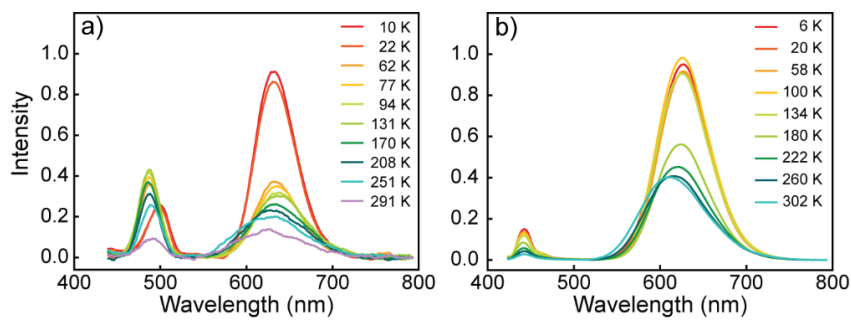

Figure S3. Temperature-dependent PL spectra of (a) Mn-doped CsPbBr$_3$ and (b) NPLs QDs dispersed in polystyrene.

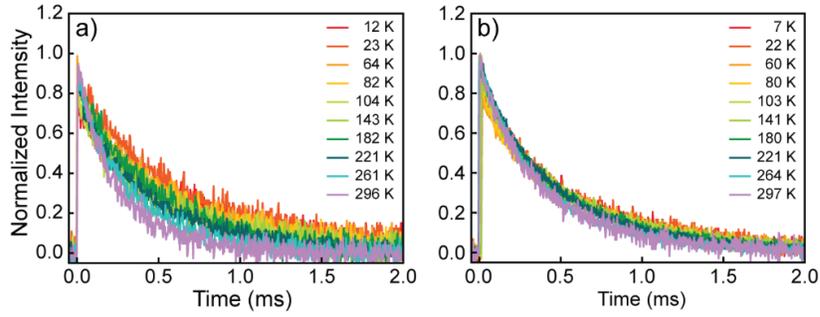

Figure S4. Time-resolved Mn$^{2+}$ PL intensities of (a) Mn-doped CsPbBr$_3$ QDs and (b) Mn-doped CsPbBr$_3$ NPLs at varying temperatures.

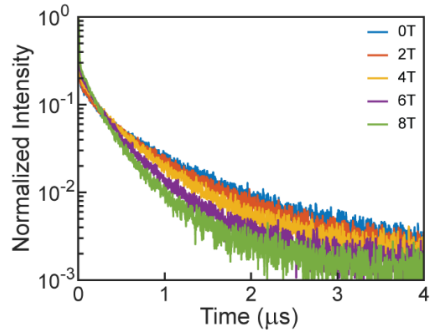

Figure S5. Magnetic field-dependent exciton PL intensity decay curves of Mn-doped CsPbBr$_3$ NPLs measured at 10 K. μs component of the PL attributed to dark exciton decays faster under the stronger magnetic field.

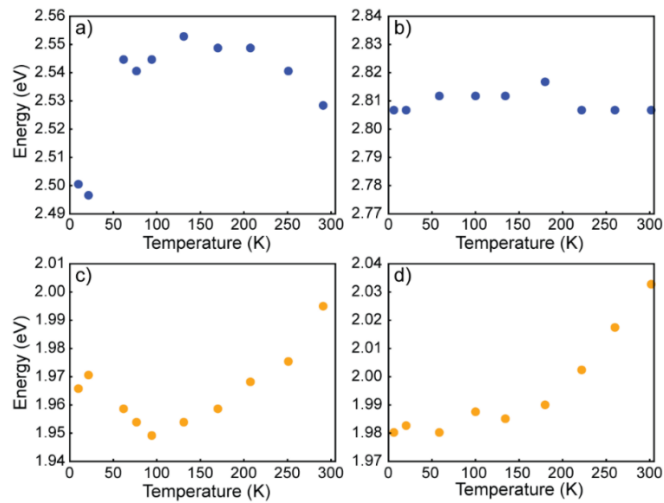

Figure S6. Temperature-dependent exciton PL energy of (a) Mn-doped CsPbBr$_3$ QDs and (b) NPLs. Temperature-dependent Mn$^{2+}$ PL energy of (c) Mn-doped CsPbBr$_3$ QDs and (d) NPLs.

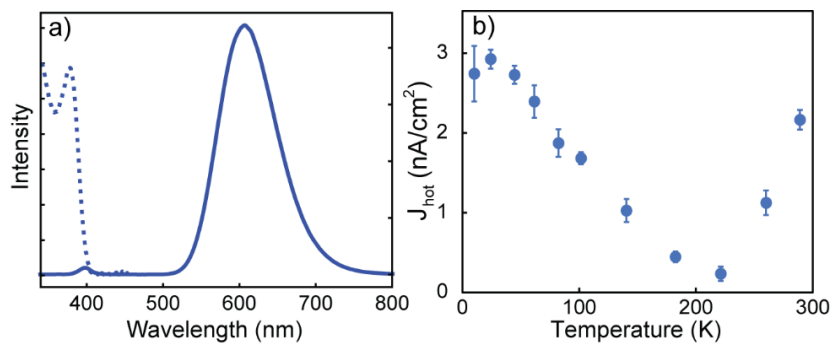

Figure S7. (a) Solution-phase absorption (dashed) and PL (solid) spectra, and (b) Temperature-dependent $J_{hot}$ of Mn-doped CsPbCl$_3$ QDs.

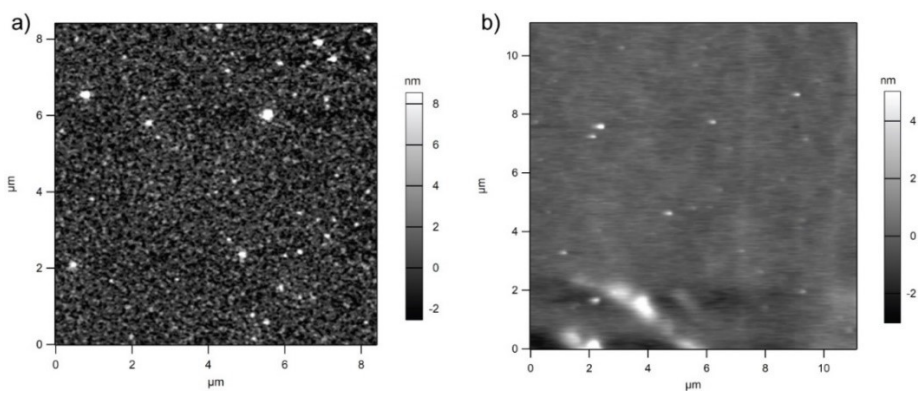

Figure S8. AFM image of (a) Mn-doped CsPbBr$_3$ QDs and (b) Mn-doped CsPbBr$_3$ NPLs spin-cast on ITO/glass substrate.

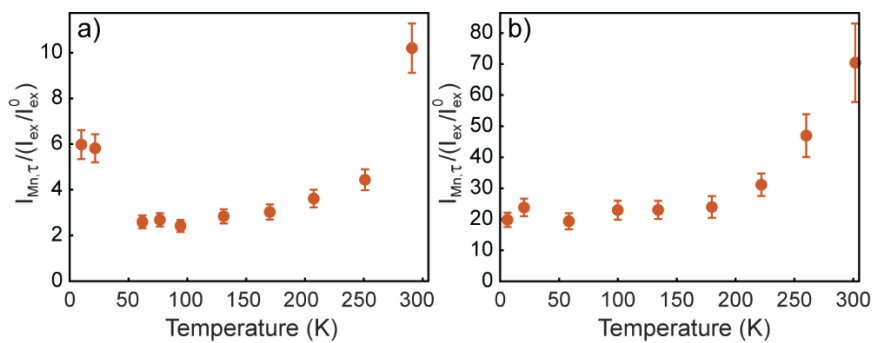

Figure. S9. $I_{Mn,\tau}/(I_{Mn}/I_{ex}^0)$ of (a) Mn-doped CsPbBr$_3$ QDs and (b) Mn-doped CsPbBr$_3$ NPLs.